\documentclass[twocolumn,showpacs,amsmath,amssymb]{revtex4}

\usepackage{enumerate}
\usepackage{amssymb} 
\usepackage{amsbsy}
\usepackage{graphicx}
\usepackage{hyperref}

\newcommand{\be}{\begin{equation}} \newcommand{\ee}{\end{equation}}
\newcommand{\bea}{\begin{eqnarray}} \newcommand{\eea}{\end{eqnarray}}
\newcommand{\el}{\nonumber \\}
\newcommand{\re}[1]{(\ref{#1})}

\newcommand{\pat}{\partial}

\newcommand{\brt}[1]{[#1]}
\newcommand{\para}{\paragraph}

\renewcommand{\a}{\alpha}
\renewcommand{\b}{\beta}
\renewcommand{\c}{\gamma}
\renewcommand{\d}{\delta}

\renewcommand{\o}{\omega}

\newcommand{\GN}{G_{\mathrm{N}}}
\newcommand{\ha}{\frac{1}{2}}

\newcommand{\rmd}{\mathrm{d}}

\newcommand{\nonum}{;\ }

\newcommand{\rhodot}{\dot{\rho}}

\newcommand{\udot}{\dot{u}}
\newcommand{\Edot}{\dot{E}}
\renewcommand{\path}{\hat{\nabla}}
\newcommand{\eps}{\mathcal{O}(\varepsilon)}
\newcommand{\one}{\mathcal{O}(1)}

\newtheorem{theorem}{Theorem}

\newcommand{\PRD}[1]{{\it Phys. Rev.} {\bf D#1}}

\newcommand{\APJ}[1]{{\it Astrophys. J.} {\bf #1}}

\newcommand{\CQG}[1]{{\it Class. Quant. Grav.} {\bf #1}}
\newcommand{\GRG}[1]{{\it Gen. Rel. Grav.} {\bf #1}}
\newcommand{\AaA}[1]{{\it Astron. \& Astrophys.} {\bf #1}}

\begin{document}

\title{On the relation between the isotropy of the CMB and the geometry of the universe}

\author{Syksy R\"{a}s\"{a}nen}
\affiliation{Universit\'e de Gen\`eve, D\'epartement de Physique Th\'eorique, \\
24 quai Ernest-Ansermet, CH-1211 Gen\`eve 4, Switzerland}

\begin{abstract}
The near-isotropy of the cosmic microwave background
(CMB) is considered to be the strongest indication for the homogeneity
and isotropy of the universe, a cornerstone of most cosmological
analysis. We derive new theorems which extend the 
Ehlers-Geren-Sachs result that an isotropic CMB
implies that the universe is either
stationary or homogeneous and isotropic, and its
generalisation to the almost isotropic case.
We discuss why the theorems do not apply to the
real universe, and why the CMB observations do
not imply that the universe would be nearly
homogeneous and isotropic.
\end{abstract}


\maketitle



\setcounter{secnumdepth}{3}

\section{Introduction} \label{sec:intro}

A central assumption in most cosmological analysis
is that the universe is homogeneous and isotropic
up to small perturbations, and thus described
by a perturbed Friedmann-Robertson-Walker (FRW) model.
The strongest observational indication of isotropy
comes from the temperature of the cosmic microwave background (CMB),
which is uniform at the $10^{-5}$ level in different directions
(excepting the dipole of $10^{-3}$) \cite{Bennett:1996, Hinshaw:2008}.

The important question is how the CMB temperature is
related to the spacetime geometry. The first step
towards constraining the geometry without assuming
that it is FRW was taken by Ehlers, Geren and Sachs,
who proved that if the only matter is a radiation fluid
with a perfectly isotropic distribution function,
the spacetime is either stationary, FRW,
or a special case with non-zero rotation
and acceleration \cite{Ehlers:1968}.
(This has been extended to arbitrary
matter with an isotropic distribution \cite{moreiso}.)
The result was generalised by Stoeger, Maartens and Ellis,
who showed that if the CMB temperature and its derivatives
are almost isotropic everywhere in an expanding
dust-dominated universe, and the observers are geodesic,
the spacetime is almost FRW \cite{SME}.

Counterexamples which violate some of the
assumptions of \cite{SME} have been presented \cite{counter}.
For cosmology, the main issue is the applicability of the
theorems to realistic models of the present-day universe,
briefly discussed in \cite{Rasanen:2008b}.
We first derive new theorems which relate the CMB temperature
to the geometry of the universe in the case of perfect
isotropy and in the case of small anisotropy,
generalising the results of \cite{Ehlers:1968, SME}.
We then discuss why neither the new nor the old theorems
apply to the real universe, and how the CMB
observations only indicate that the universe is
statistically homogeneous and isotropic, not that
it would be locally almost exactly homogeneous and isotropic
(i.e. almost FRW).

\section{The covariant formalism} \label{sec:form}


\para{The four-velocity.}

We are interested in the CMB temperature measured
by observers moving on timelike curves.
For reviews of the covariant approach we use,
see \cite{Ellis:1971, Tsagas:2007}.
The observer velocity
is denoted $u^\a$, and normalised as $u_\a u^\a=-1$.
The derivative with regard to the proper time
measured by the observers is given by $\pat_t\equiv u^\a\nabla_\a$
and also denoted by an overdot.
The tensor which projects on the space
orthogonal to $u^\a$ is $h_{\a\b} \equiv g_{\a\b} + u_\a u_\b$,
where $g_{\a\b}$ is the metric.
The derivative projected with $h_{\a\b}$
is denoted by $\path_\a\equiv h_\a^{\ \b}\nabla_\b$.

The covariant derivative of $u^\a$ can be decomposed as
\bea \label{udec}
  \nabla_\b u_\a &=& \frac{1}{3} h_{\a\b} \theta + \sigma_{\a\b} + \omega_{\a\b} -\udot_\a u_\b \ ,
\eea

\noindent where \mbox{$\theta\equiv\nabla_\a u^\a$}
is the volume expansion rate, $\sigma_{\a\b}$ is the traceless
symmetric shear tensor,
$\omega_{\a\b} \equiv \nabla_{[\b} u_{\a]} + \udot_{[\a}u_{\b]}$
is the vorticity tensor and $\udot^\a$
is the acceleration vector.
The tensors $\sigma_{\a\b}$ and $\omega_{\a\b}$ and the vector
$\udot^\a$ are orthogonal to $u^\a$.
Instead of $\omega_{\a\b}$, we can equivalently use the vorticity
vector $\omega^\a\equiv\ha\epsilon^{\a\b\c} \omega_{\b\c}$, where
$\epsilon_{\a\b\c}\equiv\eta_{\a\b\c\d} u^\d$ is the volume element
in the space orthogonal to $u^\a$, $\eta_{\a\b\c\d}$ being the
spacetime volume element.

\para{The temperature.}

In the geometrical optics approximation, light travels
on null geodesics. The null geodesic tangent vector,
identified with the photon momentum, is denoted by $k^\a$.
It satisfies $k_\a k^\a=0$ and $k^\a \nabla_\a k^\b=0$.
The energy is the projection of the momentum onto
the observer's velocity,
\bea \label{E}
  E = - u_\a k^\a \ .
\eea

\noindent The photon momentum can be decomposed into an
amplitude and the direction, and the direction
can be split into components orthogonal and parallel to $u^\a$,
\bea \label{kdec}
  k^\a = E ( u^\a + e^\a ) \ ,
\eea

\noindent where $u_\a e^\a=0$, $e_\a e^\a =1$.

In \cite{SME}, the temperature
was defined with the integrated brightness
$I\equiv\int_0^\infty \rmd E E^3 f(E,x,e)\propto T^4$,
where $f(E,x,e)$ is the phase space distribution function
of the CMB photons.
However, observations of the anisotropy of the
brightness are only made in some small energy ranges around
a few frequencies \cite{Bennett:1996, Hinshaw:2008}.
To avoid assumptions about the spectrum at unobserved
energies, rather than considering $I$, we define the
temperature using the spectrum in the observed energy range.
There the CMB distribution function
has the blackbody shape to an accuracy of $10^{-4}$
\cite{Fixsen:1996}, $f=2 (e^{E/T(x,e)}-1)^{-1}$.
We can invert $f=F(E/T(x,e))$ to obtain $E(x,e,f)=T(x,e) F^{-1}(f)$.
Neglecting interactions with matter after last
scattering, the photons are essentially collisionless,
so the distribution is conserved, $\frac{\rmd f}{\rmd v}=0$, where
$\frac{\rmd}{\rmd v}$ is the derivative along the photon flow
line in phase space. It follows that
$\frac{\rmd\ln T}{\rmd v}=\frac{\rmd\ln E}{\rmd v} =k^\a \nabla_\a \ln E$,
where the last derivative is on the spacetime manifold.
It is therefore sufficient to consider the photon
energy $E(x,e,f)$.
We assume that $E$ is an analytic function.
Decomposing $k^\a\nabla_\a E$ using \re{udec}, \re{E} and \re{kdec}
on the one hand, and writing $k^\a\nabla_\a=E ( \pat_t + e^\a \pat_\a )$
using \re{kdec} on the other, we obtain the following relation
between the energy and the spacetime geometry:
\bea \label{Egeo}
  \left( \pat_t + e^\a \pat_\a \right) \ln E &=& - \left( \frac{1}{3} \theta + \udot_\a e^\a + \sigma_{\a\b} e^\a e^\b  \right) \ .
\eea

\noindent This relation places constraints on
$\theta, \udot_\a$ and $\sigma_{\a\b}$ from the behaviour
of $E$, regardless of the matter content and the equations of motion.
For more comprehensive results about the spacetime, we
need the equations which connect these quantities
to the rest of the geometry.

\para{The evolution equations.}

We assume that matter and geometry are related
by the Einstein equation $G_{\a\b} = T_{\a\b}$,
where $G_{\a\b}$ is the Einstein tensor
and $T_{\a\b}$ is the energy-momentum tensor.
(We use units in which $8\pi\GN=1$, $\GN$ being Newton's constant.)
Without loss of generality, $T_{\a\b}$ can be decomposed as
\bea \label{emdec}
  T_{\a\b} = \rho u_\a u_\b + p h_{\a\b} + 2 q_{(\a} u_{\b)} + \pi_{\a\b} \ ,
\eea

\noindent where $\rho$ is the energy density, $p$ is the
pressure, $q_\a$ is the energy
flux and $\pi_{\a\b}$ is the anisotropic stress.
Both $q_\a$ and $\pi_{\a\b}$ are orthogonal to $u^\a$,
and $\pi_{\a\b}$ is traceless and symmetric.
The decomposition \re{emdec} can be understood as the
parametrisation of whatever tensor the Einstein tensor
is equal to, and in this sense it remains valid even in
theories where the Einstein equation does not hold.

The set of evolution equations given by the
Einstein equation, the Bianchi identity
and the Ricci identity for $u^\a$ can be conveniently written
in terms of the decompositions \re{udec} and \re{emdec}
and the electric and magnetic parts of the Weyl tensor,
$E_{\a\b} \equiv C_{\a\c\b\d} u^\c u^\d$ and
$H_{\a\b} \equiv \frac{1}{2} \epsilon_\a^{\ \c\d} C_{\c\d\b\mu} u^\mu$
\cite{Ellis:1971, Tsagas:2007}.
Let us first give those evolution equations which are
independent of the matter content and the Einstein equation,
\bea
  \label{wdot} h^\a _{\ \b} \dot{\o}^\b &=& -\frac{2}{3} \theta \o^\a +\sigma^\a_{\ \b} \o^\b - \ha \epsilon^{\a\b\c} \path_\b\udot_\c \\
  \label{wgrad} \path_\a \o^\a &=& \udot_\a \o^\a \\
  \label{H} H_{\a\b} &=& 2 \udot_{\langle \a} \o_{\b\rangle} + \path_{\langle\a} \o_{\c\rangle} + \epsilon_{\c\d\langle\a} \path^\c \sigma^\d_{\ \b\rangle} \ ,
\eea

\noindent where $\langle\rangle$ denotes the symmetric
traceless spatially projected part,
$A_{\langle\a\b\rangle}\equiv h_{(\a}^{\ \ \c} h_{\b)}^{\ \ \d} A_{\c\d} - \frac{1}{3} h_{\a\b} h^{\c\d} A_{\c\d}$.
The other equations involve the energy-momentum tensor \re{emdec},
\bea
  \label{consrho} \rhodot &=& - \theta ( \rho+p )- \path_\a q^\a - 2 \udot_\a q^\a - \sigma_{\a\b} \pi^{\a\b} \\
 \label{consp} \path_\a p &=& - \udot_\a ( \rho + p ) - h_{\a\b} \dot{q}^\b - \frac{4}{3} \theta q_\a - h_{\a\c} \path_\b \pi^{\b\c} \el
  && - \pi_{\a \b} \udot^\b - \sigma_{\a\b} q^\b + \epsilon_\a^{\ \b\c} \o_\b q_\c \\
 \label{Ray} \dot{\theta} + \frac{1}{3} \theta^2 &=& - \ha ( \rho + 3 p ) - \sigma_{\a\b} \sigma^{\a\b} + 2 \omega_\a \omega^\a \el
  && + \path_\a \udot^\a + \udot_\a \udot^\a \\
  \label{sigmadot} \dot{\sigma}_{\langle\a\b\rangle} &=& - \frac{2}{3} \theta\sigma_{\a\b} + \path_{\langle\a} \udot_{\b\rangle} + \udot_{\langle\a} \udot_{\b\rangle} - \sigma_{\c\langle\a} \sigma^{\c}_{\ \b\rangle} \el
  && - \o_{\langle\a} \o_{\b\rangle} - E_{\a\b} + \frac{1}{2} \pi_{\a\b} \\
  \label{sigmagrad} h_{\a\c} \path_\b \sigma^{\b\c} &=& \frac{2}{3} \path_\a \theta + \epsilon_\a^{\ \b\c} \left( \path_\b \o_\c + 2 \udot_\b \o_\c \right) - q_\a
\eea
\begin{widetext}
\bea
  \label{Edot} \dot{E}_{\langle\a\b\rangle} &=& - \theta E_{\a\b} + 3 \sigma_{\langle\a}^{\ \ \c} E_{\b\rangle\c} - \ha (\rho+p) \sigma_{\a\b} - \ha \sigma_{\langle\a}^{\ \ \c} \pi_{\b\rangle\c} - \ha \dot{\pi}_{\langle\a\b\rangle} - \frac{1}{6} \theta \pi_{a\b} - \ha \path_{\langle\a} q_{\b\rangle} - \udot_{\langle\a} q_{\b\rangle} \el
  && + \epsilon_{\c\d\langle\a} \left( \path^\c H_{\b\rangle}^{\ \ \d} + 2 \udot^\c H_{\b\rangle}^{\ \ \d} - \o^\c E_{\b\rangle}^{\ \ \d} - \ha \o^\c \pi_{\b\rangle}^{\ \ \d} \right) \\
  \label{Hdot} \dot{H}_{\langle\a\b\rangle} &=& - \theta H_{\a\b} + 3 \sigma_{\langle\a}^{\ \ \c} H_{\b\rangle\c} - \frac{3}{2} \o_{\langle\a} q_{\b\rangle} - \epsilon_{\c\d\langle\a} \left( \path^\c E_{\b\rangle}^{\ \ \d} - \ha \path^\c \pi_{\b\rangle}^{\ \ \d} + 2 \udot^\c E_{\b\rangle}^{\ \ \d} + \o^\c H_{\b\rangle}^{\ \ \d} + \ha q^\c \sigma_{\b\rangle}^{\ \ \d} \right) \\
  \label{Egrad} h_{\a\c} \path_\b E^{\b\c} &=& - 3 \o_\b H^{\b}_{\ \a} + \epsilon_{\a}^{\ \b\c} \left( \sigma_{\b\d} H^{\d}_{\ \c} - \frac{3}{2} \o_\b q_\c \right) + \frac{1}{3} \path_\a \rho - \ha h_{\a\c} \path_\b \pi^{\b\c} - \frac{1}{3} \theta q_\a + \ha \sigma_{\a\b} q^\b \\
  \label{Hgrad} h_{\a\c} \path_\b H^{\b\c} &=& 3 \o_\b E^{\b}_{\ \a} - \epsilon_{\a}^{\ \b\c} \sigma_{\b}^{\ \d} \left( E_{\c\d} + \frac{1}{2} \pi_{\c\d} \right) + (\rho+p) \o_\a - \ha \pi_{\a\b} \o^\b - \frac{1}{2} \epsilon_\a^{\ \b\c} \path_\b q_\c \ .
\eea
\end{widetext}

\section{Theorems} \label{sec:theorems}


\para{Isotropy theorem.}

We prove the following result about the implication
of a perfectly isotropic CMB for the spacetime geometry:
\begin{theorem}
If geodesic observers at all $x$ measure photon energy $E(x,e,f)$
that is independent of the direction $e$, the spacetime
is either stationary or obeys the conditions \re{zero1}, \re{special}.
If the anisotropic stress is zero (in particular, if the
matter is an ideal fluid), \re{zero1}, \re{special},
reduce to the statement that the spacetime is FRW.
\end{theorem}

Assuming the observers to be geodesic implies $\udot_\a=0$.
Since $E$ does not depend on $e$ and harmonic
functions of different degree are independent,
it follows from \re{Egeo} that
\bea \label{zero1}
  e^\a\pat_\a E = 0 \ , \qquad \sigma_{\a\b} = 0 \ .
\eea

\noindent Since $e^\a$ is an arbitrary direction orthogonal
to $u^\a$, $e^\a\pat_\a E=0$ is equivalent to $\path_\a E=0$,
which can be written as $\nabla_\a E=-u_\a \Edot$ by using the
definition of $h_{\a\b}$. Now there are two possibilities.
Either $\Edot=0$ or $\Edot\neq0$.

If $\Edot=0$, it follows from \re{Egeo} that $\theta=0$.
Together with $\udot_\a=0$ and $\sigma_{\a\b}=0$,
this is equivalent to $\nabla_{(\a} u_{\b)}=0$, 
so $u^\a$ is a Killing vector, and the spacetime is stationary.
The spacetime is static if and only if $u^\a$ is also
hypersurface orthogonal, which is equivalent to $\omega_{\a\b}=0$.
These results hold independently of the matter content and
the Einstein equation.

If $\Edot\neq0$, we have $\theta\neq0$.
We can write $u_\a=-\Edot^{-1}\nabla_\a E$.
This (together with $\udot_\a=0$) means
that the vorticity tensor has the form
$\omega_{\a\b}=u_{[\a} v_{\b]}$. It then follows
from $u^\a w_{\a\b}=0$ that $\omega_{\a\b}=0$.
(This is an application of Frobenius' theorem \cite{Wald:1984}.)
Given $\udot_\a=0$, the condition $\path_\a E=0$ implies
$\path_\a \Edot=0$, so from \re{Egeo} we get $\path_\a\theta=0$.
From \re{H} we have $H_{\a\b}=0$.
These results hold independently of the matter content and
the Einstein equation.
Assuming that the Einstein equation holds,
the evolution equations \re{wdot}--\re{Hgrad}
reduce to the following:
\bea \label{special}
 && \o_{\a\b} = 0 \ , \qquad \path_\a\theta = 0 \ , \qquad H_{\a\b} = 0 \el
  && \rhodot + \theta ( \rho + p ) = 0 \ , \qquad \dot{\theta} + \frac{1}{3} \theta^2 = - \ha ( \rho + 3 p ) \el
  && \path_\a ( \rho + 3 p ) = 0 \el
  && \path_\b \pi^\b_{\ \a} = - \path_\a p \ , \qquad \dot{\pi}_{\a\b} + \frac{2}{3} \theta \pi_{\a\b} = 0 \el
  && E_{\a\b} = \frac{1}{2} \pi_{\a\b} \ , \qquad q_\a = 0 \ .
\eea

\noindent It follows that if the anisotropic stress is zero,
the spacetime is FRW. In particular, this is true for an ideal
fluid. The result that for an ideal fluid the conditions
$\udot_\a=0, \sigma_{\a\b}=\omega_{\a\b}=0$ imply
that the spacetime is FRW is well-known \cite{Ellis:1971}.
Further, for an ideal fluid the conditions $\udot_\a=0, \sigma_{\a\b}=0$
imply that $\theta\omega_{\a\b}=0$ \cite{Senovilla:1997},
so the spacetime is either stationary or FRW (or both).

Note that we have made no assumptions about the
photon distribution function outside the observed
frequency range or about the distribution of
matter components other than the CMB photons,
in contrast to \cite{Ehlers:1968, moreiso, SME}.


\para{'Almost isotropy' theorem.}

We generalise the previous theorem to the case
of small anisotropy:
\begin{theorem}
If observers at all $x$ measure photon energy $E(x,e,f)$
such that the first three harmonic moments of
$(\pat_t + e^\a \pat_\a)\ln E$ and their derivatives
are independent of the direction $e$ to $\eps$
(where $\varepsilon\ll1$ is a constant),
and the matter is an ideal fluid to $\eps$, the
spacetime is, to $\eps$, either stationary or FRW.
\end{theorem}

\noindent Note that the observers are not assumed to
be geodesic. The detailed conditions on $(\pat_t + e^\a \pat_\a) \ln E$
and the matter content are given in the proof below.

When $E$ is not perfectly isotropic, knowing the anisotropy
of $E$ is not sufficient to place constraints on the local
geometry, because the relation \re{Egeo} involves the
derivatives of $E$.
In fact, we only need assumptions about the derivatives of
$\ln E$. We expand their direction dependence in covariant harmonics:
\bea
  && (\pat_t + e^\a \pat_\a) \ln E(x,e,f) \equiv A = \bar{A}(x) + A_\a(x) e^\a \el
  && + A_{\a\b}(x) e^\a e^\b + \sum_{n=3}^{\infty} A_{\a_1\dots\a_n}(x) e^{\a_1} \dots e^{\a_n} \ .
\eea

\noindent Since \re{Egeo} involves at most two $e^\a$ on the
right-hand side, we do not need moments of $A$
higher than two, corresponding to the dipole
and the quadrupole (of the derivatives, not the energy itself).
Decomposed into harmonic moments, the relation \re{Egeo} reads
\bea \label{Egeodec}
   \bar{A} = - \frac{1}{3} \theta \ , \qquad
  A_\a = - \udot_\a \ , \qquad
   A_{\a\b} = - \sigma_{\a\b} \ .
\eea

If we assume that the anisotropy of $A$ is small,
we have to say with respect to which scale this holds,
since $A$ is a dimensional quantity.
We introduce a timescale $L$, and assume that
$A_\a, A_{\a\b}\lesssim\eps L^{-1}$
where $\varepsilon\ll1$ is a fixed constant.
(The scale $L$ need not be constant; in fact,
when $\bar{A}$ is not also small, a natural choice
of $L$ is the local Hubble time
$3 \theta^{-1}=-\bar{A}^{-1}$.)
From \re{Egeodec} we have
\bea \label{zero2}
  \udot_\a \lesssim \eps L^{-1} \ , \qquad
  \sigma_{\a\b} \lesssim \eps L^{-1} \ .
\eea

\noindent We consider two possibilities, either
$\bar{A}\lesssim\eps L^{-1}$ or $\bar{A}$ is
of the same order as $L^{-1}$.

If $\bar{A}\lesssim\eps L^{-1}$, we have from \re{Egeodec}
the result $\theta\lesssim\eps L^{-1}$, which combined with
\re{zero2} means $\nabla_{(\a} u_{\b)}\lesssim\eps L^{-1}$.
So $u^\a$ is an 'almost Killing vector', and the spacetime
is 'almost stationary' when viewed on timescales
of $L$ or shorter.
If the vorticity is also $\eps L^{-1}$, the
spacetime is 'almost static'.

If $\bar{A}=\one L^{-1}$, we have
$\theta=\one L^{-1}$.
Analogously to the exactly isotropic case, we write
$u_\a=-\dot{\bar{A}}^{-1}\nabla_\a\bar{A} + \dot{\bar{A}}^{-1}\path_\a\bar{A}$.
Assuming that $\nabla_\a\path_\b\bar{A}\lesssim\eps L^{-3}$
and $\dot{\bar{A}}=\one L^{-2}$, it follows that
$\omega_{\a\b}\lesssim\eps L^{-1}$.
The conditions
$\udot_\a, \sigma_{\a\b}, \omega_{\a\b}\lesssim\eps L^{-1}$
do not necessarily imply that the spacetime is FRW to
$\eps$ even if the matter is an ideal fluid, because the evolution
equations \re{wdot}--\re{Hgrad} involve their derivatives,
which may be large. We assume that
$\path_\a\bar{A}, \path_\a\nabla_\b\path_\c\bar{A}, \path_\a A_\b, \path_\a\path_\b A_\c, \nabla_\a A_{\b\c}$,
as well as $\nabla_\a q_\b, \pi_{\a\b}, \nabla_\a \pi_{\b\c}$,
are all $\lesssim\eps$.
Then \re{wdot}--\re{Hgrad}
reduce at leading order to the FRW equations,
and spatial derivatives and Weyl tensor components
are $\eps$. In this sense, the spacetime is FRW to $\eps$.
(The assumption about $\nabla_\a q_\b$ could be replaced with further
assumptions about the derivatives of $\bar{A}, A_\a$ and $A_{\a\b}$.)

\section{Discussion} \label{sec:disc}

\para{The real universe.}

Theorem 1 follows from the directional
distribution of photon energies, which is observable.
However, since the observed CMB is not perfectly
isotropic, the theorem is not relevant
for the real universe.
In contrast, theorem 2 relies on assumptions
about the derivatives of the energy, which are not directly observed.
It would be more accurate to say that the 'almost isotropy'
theorem characterises the relation between the geometry and
the CMB in nearly stationary or nearly FRW universes,
rather than that it places constraints on the geometry from CMB
observations.
The same is true for the theorem of \cite{SME}, where
it was assumed that derivatives of the temperature
anisotropy of up to third order are small.
The smallness of the derivatives may seem to be just
a technical assumption, and in \cite{SME} it was justified
with the Copernican principle.
However, the derivatives are related to
the local geometry, and large local variations
are not in contradiction with the assumption
that our position is typical.
Extending the Copernican principle to mean
that local variations are small is equivalent to
assuming that the universe is almost FRW, and this can be done
for the geometry without going via the CMB.

In the real universe, there are large spatial
variations in the geometry. For example, the volume
expansion rate $\theta$ varies by a factor of order unity
between underdense and overdense regions.
Reading the relation \re{Egeo} between the
energy and the geometry the other way than we have
done so far shows that the derivatives
of $E$ are therefore large.
This does not imply that the energy would have
large spatial variations and thus be anisotropic
(otherwise there would be no
need to make assumptions about the
derivatives of $E$ for the 'almost isotropy' theorem).
This can be made transparent by
integrating \re{Egeo} to obtain
$E(x,e,f) = E_{\mathrm{s}} \exp\left( - \int_{\mathrm{s}}^{\mathrm{o}} \rmd\eta \left[ \frac{1}{3}\theta + \udot_\a e^\a + \sigma_{\a\b} e^\a e^\b \right]\right)$,
%
where the integral is from a source to the 
observer along the null geodesic which has the tangent vector $e^\a$.
Here $\eta$ is a parameter along the geodesic
defined by $\frac{\rmd}{\rmd\eta}\equiv\pat_t +e^\a\pat_\a$.
The CMB photons start from the last
scattering surface, and $E_{\mathrm{s}}$ is given by the
distribution $f$ and the constant decoupling temperature.
This relation shows how the energy itself, instead of
its derivatives, is related to the geometry.
As long as the spatial variation in the geometry
is uncorrelated over long distances, and the coherence scale
is small compared to the magnitude of the derivatives,
the anisotropy in $E$ is small \cite{Rasanen:2008b}.

\para{Conclusion.}

A perfectly isotropic CMB seen by geodesic observers
implies that the spacetime is stationary or FRW
(or has the restricted form \re{zero1}, \re{special}),
but an almost isotropic CMB implies that the
universe is almost stationary or almost FRW only with additional
assumptions about the derivatives of the CMB temperature.
These assumptions do not hold in the real universe.
The observed isotropy of the CMB, coupled
with the Copernican assumption and analyticity,
indicates statistical homogeneity and isotropy,
but not local homogeneity and isotropy.
This is an important distinction, because a universe which
is only statistically homogeneous and isotropic is not
described by the FRW metric, and its
expansion is not determined by the FRW equations.
In particular, the assumption that the universe is FRW is crucial
in deducing the existence of dark energy or modified gravity.
Dropping this assumption, it may be possible to
explain the observed deviation in the expansion
of the universe at late times from the matter-dominated
FRW prediction without new physics
\cite{Rasanen:2006b, Rasanen:2008a, Rasanen:2008b}.

\para{Acknowledgments.}

I thank Ruth Durrer for helpful discussions
and comments on the manuscript, and Roy Maartens
for comments and for pointing out an error in the
first version of the paper.

\setcounter{secnumdepth}{0}


\end{document}